\documentclass[iicol,sn-mathphys-num]{sn-jnl}% Math and Physical Sciences Numbered Reference Style 
\usepackage{graphicx}%
\usepackage{multirow}%
\usepackage{amsmath,amssymb,amsfonts}%
\usepackage{amsthm}%
\usepackage{mathrsfs}%
\usepackage[title]{appendix}%
\usepackage{multirow}
\usepackage{xcolor}%
\usepackage{soul}
\usepackage{textcomp}%
\usepackage{manyfoot}%
\usepackage{booktabs}%
\usepackage{graphicx}
\usepackage{textcomp}
\usepackage{algorithm}%
\usepackage{algorithmicx}%
\usepackage{algpseudocode}%
\usepackage{listings}%
\theoremstyle{thmstyleone}%
\theoremstyle{thmstyletwo}%

\theoremstyle{thmstylethree}%

\begin{document}

\title[Article Title]{Leveraging Vision Transformers for Enhanced Classification of Emotions using ECG Signals}


\author*[1,2]{\fnm{Pubudu L.} \sur{Indrasiri}}\email{pranpatidewage@deakin.edu.au}

\author[2,3]{\fnm{Bipasha} \sur{Kashyap}}\email{b.kashyap@deakin.edu.au}
\equalcont{These authors contributed equally to this work.}

\author[1,2]{\fnm{Pubudu N.} \sur{Pathirana}}\email{pubudu.pathirana@deakin.edu.au}
\equalcont{These authors contributed equally to this work.}

\affil*[1,2,3]{\orgdiv{School of Engineering}, \orgname{, Deakin University}, \orgaddress{\street{75, Pigdons Rd}, \city{Waurn Ponds}, \postcode{3216}, \state{VIC}, \country{Australia}}}


\abstract{Biomedical signals provide insights into various conditions affecting the human body. Beyond diagnostic capabilities, these signals offer a deeper understanding of how specific organs respond to an individual's emotions and feelings. For instance, ECG data can reveal changes in heart rate variability linked to emotional arousal, stress levels, and autonomic nervous system activity. This data offers a window into the physiological basis of our emotional states. Recent advancements in the field diverge from conventional approaches by leveraging the power of advanced transformer architectures, which surpass traditional machine learning and deep learning methods. We begin by assessing the effectiveness of the Vision Transformer (ViT), a forefront model in image classification, for identifying emotions in imaged ECGs. Following this, we present and evaluate an improved version of ViT, integrating both CNN and SE blocks, aiming to bolster performance on imaged ECGs associated with emotion detection. Our method unfolds in two critical phases: first, we apply advanced preprocessing techniques for signal purification and converting signals into interpretable images using continuous wavelet transform and power spectral density analysis; second, we unveil a performance-boosted vision transformer architecture, cleverly enhanced with convolutional neural network components, to adeptly tackle the challenges of emotion recognition.  Our methodology's robustness and innovation were thoroughly tested using ECG data from the YAAD and DREAMER datasets, leading to remarkable outcomes. For the YAAD dataset, our approach outperformed existing state-of-the-art methods in classifying seven unique emotional states, as well as in valence and arousal classification. Similarly, in the DREAMER dataset, our method excelled in distinguishing between valence, arousal and dominance, surpassing current leading techniques.}

\keywords{Biomedical signals, vision transformers, ECG, wavelet transform, power spectral density, emotion recognition}



\maketitle

\section{Introduction}\label{sec1}

Automatic detection of emotions plays a pivotal role in affective computing, finding successful integration across diverse fields including multimedia applications \cite{mariappan2012facefetch}, biopsychosocial healthcare systems [3], and human-computer interaction (HCI) [4]. Advancements in wearable technology have significantly boosted research into multisensory data acquisition and analysis for emotion detection [5], [6]. Multisensory or multimodal data, gathered using various sensors across different modalities, encompass a wide range of inputs such as images of facial expressions, vocal and speech patterns, and physiological signals.

In the landscape of emotion recognition, bio-signal-based
methods \cite{awan2022ensemble}, \cite{bulagang2020review}, \cite{soleymani2011multimodal}, \cite{subramanian2016ascertain} are highly accurate and are not
susceptible to being masked, unlike other methods such as
facial emotion recognition and speech analysis \cite{kim2008emotion}. Advances in Human-Computer Interaction (HCI) technologies have led to the creation of sophisticated multimodal databases for emotion recognition \cite{miranda2018amigos}, \cite{katsigiannis2017dreamer}, \cite{dar2022yaad}. These databases encompass a wide array of physiological signals, aiming to construct a detailed emotional profile that includes affect (the experience of feeling or emotion), valence (the positive or negative quality of an emotion) and arousal (the level of alertness or excitement). Central to these collections are signals such as electroencephalography (EEG), facial electromyography (EMG), electrocardiography (ECG), and galvanic skin response (GSR). Each modality contributes unique dimensions to emotion recognition, enabling more nuanced and precise interpretations of affective states. Table \ref{my-label} delineates the salient features of several prominent datasets in this domain, providing a comparative overview of their composition and the physiological signals they encompass. These databases are typically generated under controlled laboratory conditions, where participants' emotions are induced through the viewing of emotionally charged video content.

\begin{table*}[ht]
\centering
\caption{Summary of Emotion Recognition Datasets}
\label{tbl:emotion_datasets}
\renewcommand{\arraystretch}{1.3}
\setlength{\tabcolsep}{4pt}
\footnotesize  % Reduced font size for better fit
\begin{tabular*}{\textwidth}{@{\extracolsep{\fill}}llp{3 cm}p{2.5 cm}p{3 cm}p{3 cm}}
\toprule
\textbf{Dataset} & \textbf{Partic.} & \textbf{Modalities} & \textbf{Videos \& Duration} & \textbf{Annotations} & \textbf{Use Case} \\
\midrule
AMIGOS \cite{miranda2018amigos} & 40 & EEG, ECG, GSR, Video, Facial Exp. & 20 (50--150 sec) & Self-assess., Valence, Arousal & Emotion Rec., Multimodal Interaction \\
DEAP \cite{koelstra2011deap} & 32 & EEG, ECG, GSR, EMG, Video & 40 (60 sec) & Valence, Arousal, Dominance, Liking & Physio. Signal Analysis \\
DREAMER \cite{katsigiannis2017dreamer} & 23 & EEG, ECG & 18 (67--394 sec) & Valence, Arousal & EEG, ECG-based Emotion Rec. \\
MAHNOB-HCI \cite{soleymani2011multimodal} & 27 & EEG, Peripheral Signals, Video & 20 (34--117 sec) & Emotion Labels, Valence, Arousal & Affective Comp., HCI \\
YAAD \cite{dar2022yaad} & 25 & ECG, GSR & 21 (39 sec) & Emotion Labels, Valence, Arousal & Complex Emotion Rec. \\
\bottomrule
\end{tabular*}
\end{table*}

Notwithstanding the increased accuracy, the deployment of multiple sensors has occasionally resulted in user discomfort or dissatisfaction \cite{nikolova2018ecg}. This underscores the necessity of balancing technical precision with practical usability in the design of emotion recognition systems. Within the spectrum of biosignal sensors, the electrocardiogram (ECG) emerges as a predominant choice \cite{nikolova2018ecg}, \cite{bexton1986diurnal}. Its ubiquity is grounded in the reliability of ECG signals, which are notably robust against noise and their proven correlation with emotional state.

Conventional emotion recognition machine learning techniques, such as Gaussian Naive Bayes, Support Vector Machines, k-Nearest Neighbors, and Random Forests \cite{subramanian2016ascertain}, \cite{miranda2018amigos}, \cite{hsu2017automatic},\cite{shu2020wearable}, rely on expert-driven manual selection of temporal and spectral features. While these methods of feature extraction are intricate, they are hindered by suboptimal predictive accuracy \cite{dissanayake2019ensemble}. Addressing these challenges, emotion recognition has evolved, with deep learning models now at the forefront \cite{dessai2023emotion},\cite{fan2023new}, \cite{santamaria2018using}, harnessing physiological signals for enhanced performance. The predominant deep learning strategies encompass unimodal and multimodal tasks, with the 1D-CNN \cite{santamaria2018using} and hybrid 2D-CNN-LSTM \cite{dar2020cnn} frameworks being particularly prevalent. In these CNN-based methods, physiological signals are transformed into visual representations through spectrograms \cite{rahim2019emotion} and scalograms \cite{dessai2023emotion} through wavelet transforms before CNN processing.

However, conventional convolutional neural network (CNN) methodologies exhibit inherent limitations, particularly in processing complex data with long-range dependencies. This shortcoming is crucial in the context of emotion classification using ECG data, where spatial relationships across the data significantly influence the identification of emotional states. In contrast, Vision Transformers (ViTs) \cite{dosovitskiy2020image} effectively capture these long-range dependencies through their self-attention mechanism, allowing for a comprehensive assessment of the entire input space, a critical feature for interpreting ECG images. ViTs dynamically focus on salient features across the dataset, irrespective of their spatial location, adapting effectively to the nuanced demands of emotion recognition from biomedical signals. These architectures have demonstrated utility across multimodal inputs including text, visuals, audio, and physiological data \cite{tsai2019multimodal}, \cite{wu2019attending}, \cite{huang2020multimodal}, \cite{cai2021multimodal}, \cite{chien2021self}, and have been extended to general time-series analysis \cite{wu2020deep}. Notably, the study by Arjun et al. \cite{arjun2021introducing} adapted the Vision Transformer for EEG signal interpretation, employing continuous wavelet transform to create image-based signal inputs, demonstrating the versatility and effectiveness of ViTs in signal processing. The integration of ViTs into emotion recognition represents a transformative step towards more accurate and responsive healthcare diagnostics, potentially enhancing patient monitoring and treatment strategies. 

In this study, we present a groundbreaking framework that significantly advances emotion detection from ECG data by leveraging an optimized Vision Transformer architecture. The proposed approach involves transforming ECG signals into a composite three-channel image through Continuous Wavelet Transform (CWT) and Power Spectral Density (PSD) and the model builds on the ViT architecture and introduces
a CNN block which integrates with squeeze and excitation blocks that are used to create an embedding of the full input image, which is then iteratively fed to each Transformer encoder layer by concatenating the image embedding to the output
of each transformer encoder layer. Rigorously validated against the ECG component of the YAAD and DREAMER datasets, our methodology not only pioneers the use of Vision Transformers for unimodal physiological signal analysis but also sets a new benchmark in accuracy, surpassing existing state-of-the-art methods.

\section{Related Works}

In this section, we explore the corpus of related research encompassing multimodal emotion detection, ECG-centric approaches to emotion discernment, and the application of deep learning methodologies within the realm of emotion detection.

\subsection{Multimodal Emotion Detection}

A novel ensemble learning method integrating EEG, ECG, and GSR signals achieved an impressive 94.5\% accuracy on the AMIGOS dataset, demonstrating the potential of ensemble approaches in this domain \cite{awan2022ensemble}. Comprehensive reviews provide overviews of emotion classification techniques using ECG and GSR signals, delineating the evolution and effectiveness of these methods \cite{bulagang2020review}, \cite{dessai2023emotion}. Practical applications using SVM classifiers on ECG and GSR data have shown varying degrees of success; studies with the MAHNOB database reported accuracies around 46\% for Arousal and 45.5\% for Valence \cite{soleymani2011multimodal}, while another study using the ASCERTAIN database reported slightly higher accuracies \cite{subramanian2016ascertain}. Additionally, innovative approaches employing deep learning and multimodal models to utilize EEG alongside peripheral physiological signals mark a significant shift towards more sophisticated, accurate, and reliable emotion detection systems \cite{santamaria2018using}, \cite{zhao2019multimodal}.
 
 \subsection{ECG based emotion detection}
 Sayed Ismail et al. \cite{ismail2021evaluation}. converted ECG data from the DREAMER database into images and obtained an accuracy of 63\% for Valence and an accuracy of 58\% for Arousal. They further obtained an accuracy of 79\% for Valence and an accuracy of 69\% for Arousal for numerical ECG data using the SVM classifier, proving that ECG numerical data give better classification accuracy than ECG images. The study \cite{bulagang2021multiclass} used a virtual reality headset to allow subjects to view 360-degree video stimuli. They recorded ECG signals from 20 participants using the Empatica E4 wristband. Inter-subject classification achieved 46.7\% accuracy for SVM, 42.9\% for KNN, and 43.3\% for Random Forest. A valence and arousal accuracy of 62.3\% was obtained for ECG signals from the DREAMER for emotion classification \cite{katsigiannis2017dreamer}. Miranda-Correa et al. \cite{miranda2018amigos} obtained classification accuracies of 59.7\% for Valence and 58.4\% for Arousal using ECG data. The study \cite{fan2023new} developed a deep convolutional neural network with attention mechanisms, achieving improved emotion recognition accuracies using ECG data: 96.5\% on the WESAD dataset, 83.6\% for arousal and 84.2\% for valence on the DREAMER dataset, and 68.0\% for arousal and 64.5\% for valence on the ASCERTAIN dataset. These results demonstrate the model's effectiveness across multiple datasets. In the study \cite{khan2022evaluation}, Extra Trees and Multi-Layer Perceptron (MLP) algorithms were assessed for ECG-based emotion recognition. On the DREAMER dataset, it excelled in valence prediction (74.6\%) and MLP in arousal prediction (74.6\%). The study's \cite{sarkar2020self} self-supervised model for ECG-based emotion recognition achieved accuracies of 79.6\% and 78.3\% for arousal and valence in AMIGOS, 77.1\% and 74.9\% in DREAMER, 95.0\% in WESAD, and 92.6\%, 93.8\%, and 90.2\% for arousal, valence, and stress in SWELL, demonstrating robust performance across multiple datasets.

\subsection{Deep Learning for Emotion detection}

A notable strategy, as detailed by \cite{santamaria2018using}, involves deploying a 1D Convolutional Neural Network (CNN) for feature extraction and subsequently using a fully connected network (FCN) for emotion classification. An innovative variation by Harper and Southern [32] integrates a long-short-term memory (LSTM) network with a 1D-CNN for a combined approach. In a different tactic, Siddharth et al. [33] transform signals into images via spectrograms [34], employing a 2D-CNN for extracting features, and an extreme learning machine [35] for the classification phase, showcasing the versatility of deep learning in advancing emotion recognition research. The study \cite{dessai2023emotion} explores emotion classification with CWT features and various CNN models, achieving high accuracy up to 99.19\%. The study \cite{fan2023new} presents a new deep convolutional neural network incorporating attentional mechanisms for ECG emotion recognition.

In terms of transformer approaches, the study \cite{wu2023transformer} presents a self-supervised learning framework using transformers for effective fusion of multimodal data in wearable emotion recognition. The study \cite{siriwardhana2020multimodal} introduces a Transformer-based fusion mechanism for self-supervised multimodal emotion recognition.

\section{Motivations and Contributions}
The field of emotion detection from biosignals is increasingly gravitating towards computer vision techniques, with Vision Transformers (ViTs) emerging as a potent tool outperforming Convolutional Neural Networks (CNNs) in specific scenarios. This shift highlights a promising yet underexplored avenue for ECG-based emotion detection, where the unique capabilities of ViTs have not yet been applied. Given the sparse research focusing solely on ECG signals for emotion recognition. In this regard, our main contributions are as follows:
\begin{itemize}
  \item We propose a performance-enhanced Vision Transformer architecture tailored for ECG-based emotion detection, leveraging spatial-temporal ECG signal characteristics.
  \item A novel technique for generating three-channel images from ECG signals is introduced, enabling the application of ViTs for improved feature extraction.
  \item The proposed model is validated against the YAAD and DREAMER datasets, demonstrating superior performance over existing methods and establishing a new benchmark in the field.
\end{itemize}

\section{Materials and Methods}
Our proposed framework encompasses three distinct phases: 1) Signal preprocessing, 2) Conversion of signals into images, and 3) Application of the images to a performance-enhanced Vision Transformer model.

\subsection{Dataset Descriptions}

\subsubsection{\textbf{YAAD}}The YAAD dataset, presented by Dar et al. \cite{dar2022yaad}, contains different biosignals of subjects exposed to stimulus of seven different emotions through video visualization. The YAAD dataset is composed of two subsets: a single-modal subset which contains ECG signals from 13 subjects up to three rounds for some of them, resulting in 154 single-channel samples; a multi-modal subset which contains 3 rounds of both ECG and GSR signals from another 12 different subjects, resulting in 252 two-channel samples. ECG signals were acquired at a sampling frequency of 128 Hz and have a duration of 39 s. On the contrary, GSR samples have a sampling frequency of 256 Hz and the same duration.

\subsubsection{\textbf{DREAMER}}
The DREAMER data set is a multimodal emotion data set developed by Katsigiannis and Ramzan [46]. The DREAMER data set consists of EEG and ECG signals from 23 subjects (14 males and 9 females). The participants watched 18 film clips to elicit nine different emotions. After watching a clip, the self-assessment manikins were used to acquire assessments of valence, arousal, and dominance.

\subsection{Signal Preprocessing}
Given \(s[n]\) as the raw ECG signal, where \(n\) represents the discrete time index, and \(f_s\) as the sampling frequency, which in this scenario is \(f_s = 128\) Hz.

\subsubsection{Baseline Removel}
Initially, the study \cite{dar2022yaad} highlighted that the stimulus initiation occurs after the initial five-second interval. Thus, the baseline period in the samples is calculated by \( \text{Baseline}_{\text{samples}} = BW \times f_s \), where \(BW\) is the baseline duration in seconds, in this scenario 5 seconds. Then, the baseline is calculated as the average value of the signal over the baseline window. If we let $b$ be the baseline, then it can be calculated as:
\begin{equation}
b = \frac{1}{BW_{\text{samples}}} \sum_{n=0}^{BW_{\text{samples}} - 1} s[n]
\end{equation}
Finally, the signal with the baseline removed, \(s_{\text{br}}[n]\), is then calculated by subtracting the baseline from the original signal for each sample, expressed as \(s_{\text{br}}[n] = s[n] - b\), and pass to the filtering process.\\

\subsubsection{ECG Filtering}
The baseline removed signal ($s_{br}$), undergoes a pre-filtering process using a second-order band-pass Butterworth filter. This filter, with cutoff frequencies set at 0.5 Hz and 15 Hz, is applied to mitigate the impact of environmental noise and muscle movements. This ensures the purity of the signal and enhances its suitability for subsequent analysis.
\begin{figure*}
	\centering
	\includegraphics[width=1\linewidth]{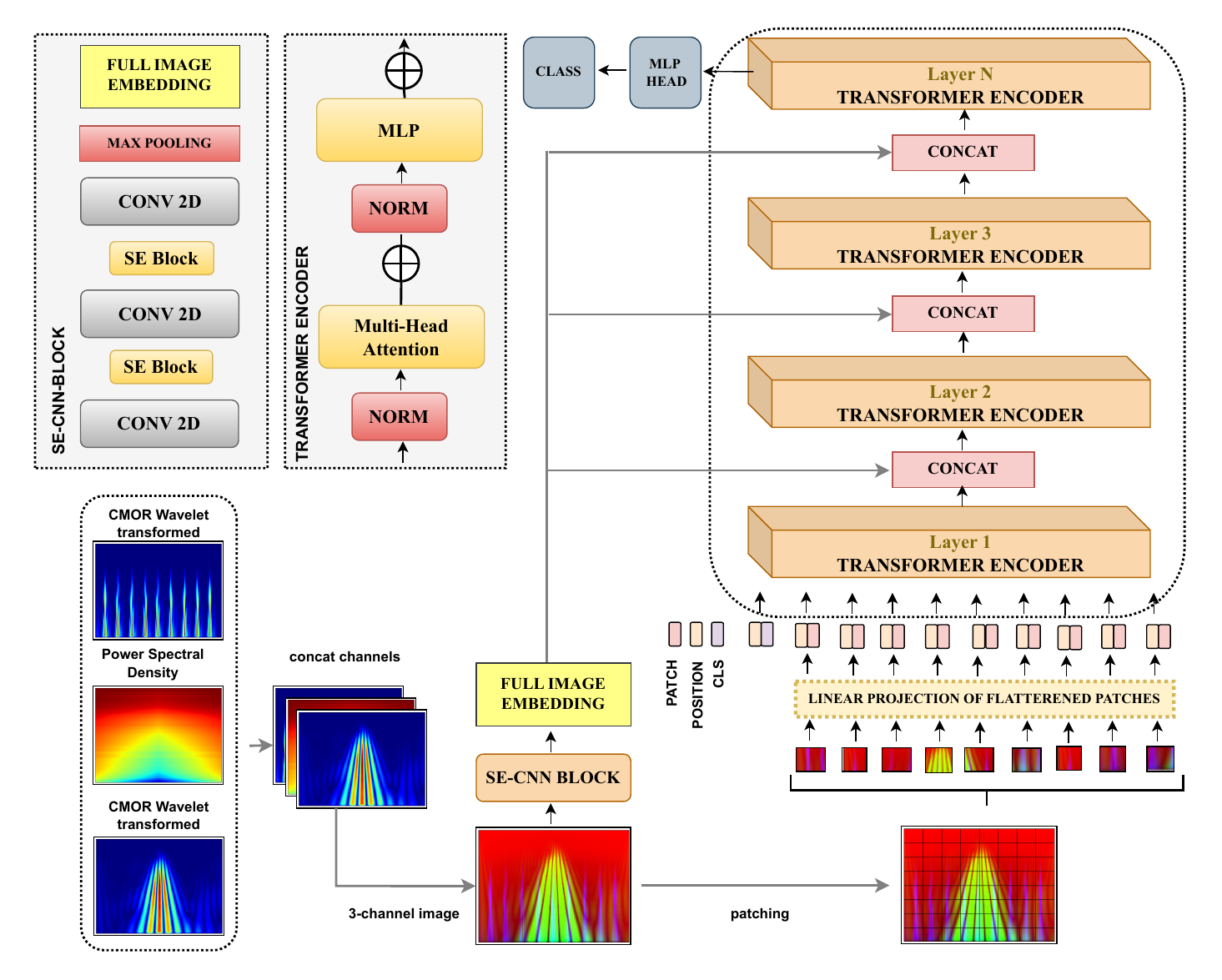}
	\caption{Proposed architecture for ECG data classification using vision transformers.}
	\label{f01}
	\vspace{-0.1in}
\end{figure*}

\subsection{ECG Signal Segmentation}
In our study, segmentation involved isolating a full cycle of the ECG signal from the overall waveform. To accomplish this, we utilized the PeakUtils Python library to identify the R-peaks within the filtered ECG signal. The parameter 'thres=0.5' specifies the relative threshold for detecting peaks in the signal. A peak is identified if its amplitude is at least 50\% of the maximum amplitude observed in the signal after filtering.  These peaks served as reference points for segmentation. For each detected R-peak, we segmented the signal by extracting 100 samples to the left and 100 samples to the right of the peak, resulting in segments of a fixed size of 200 samples each. This method ensured consistent segmentation across the ECG dataset for analysis. All the signal processing steps are visualized in Fig. \ref{f01}.

\begin{figure}
	\centering
	\includegraphics[width=.95\linewidth]{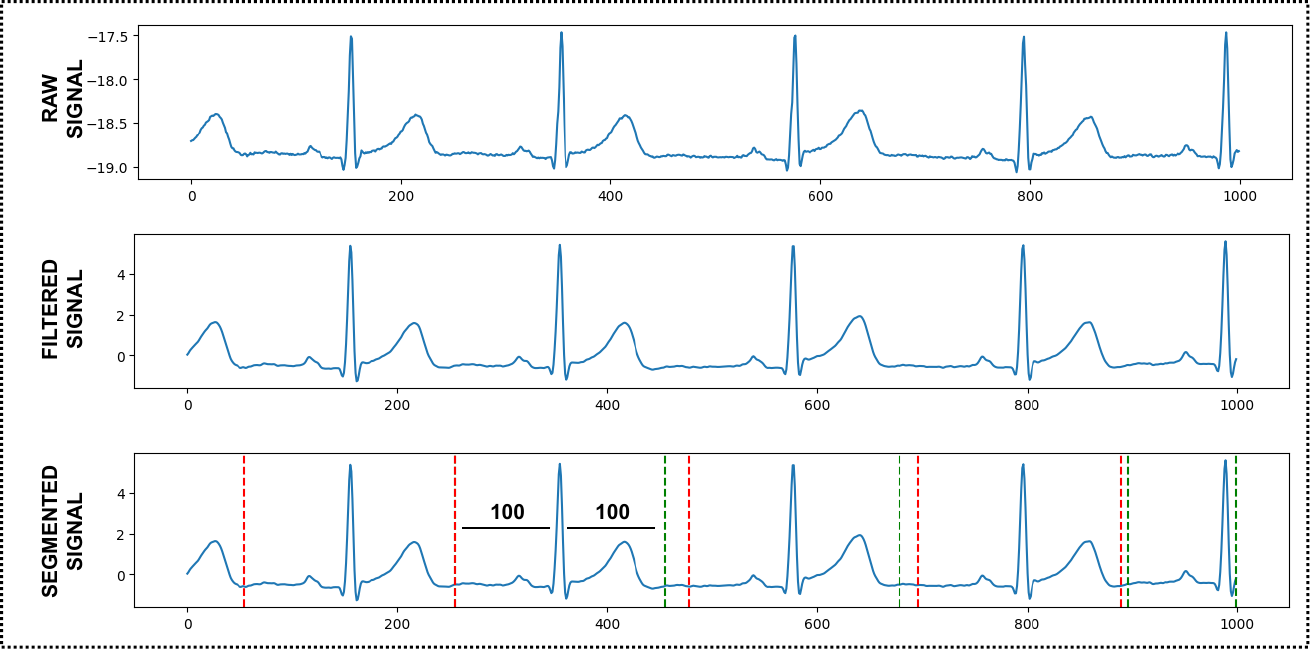}
	\caption{Signal Processing steps.}
	\label{f01}
	\vspace{-0.1in}
\end{figure}

\subsection{ECG Image Encoding}
Given our objective to leverage Vision Transformers for analysis, it is imperative to transform the signal data into a visual format. We opted for the wavelet transform approach due to its dual capacity to encapsulate information pertinent to both time and frequency domains. This choice aligns with the intrinsic architecture of Vision Transformers, which necessitates input in an image format, thereby enabling a comprehensive analysis that integrates temporal dynamics with frequency characteristics. 

\subsection{Continuous Wavelet Transform}
CWT is a powerful tool for time-frequency analysis. Unlike Fourier Transform, which only provides frequency information, CWT maintains both time and frequency information. This makes CWT particularly suited for analyzing signals where the frequency components vary over time, as is often the case with ECG and GSR signals.

For our analysis, we employed the complex Morlet wavelet, also known as the Gabor wavelet, with 50 band-pass filter banks. This wavelet is renowned for its equal variance in both time and frequency domains, offering a balanced analysis framework. This selection was made to take advantage of the Morlet wavelet's capacity for precise time-frequency localization, essential for capturing the nuanced dynamics of ECG and GSR signals.

\subsection{Power Spectral density}
The PSD is a common way to analyze the frequency content of a signal, providing insights into the power distribution across various frequency bands. This transformation is particularly useful in understanding the underlying physiological processes and detecting abnormalities in ECG signals.

Welch's method (scipy.signal.welch) divides the signal into overlapping segments, applies a window to each segment, computes the periodogram for each segment, and then averages these periodograms to estimate the PSD. Welch's method can be applied to the entire signal without prior segmentation by the user, as the method itself handles the segmentation internally.

\subsection{RGB Image formation}

In terms of a single participant, we meticulously applied the wavelet transform to each segmented portion of the signal, thereby producing multiple 2D representations for the individual. Subsequent to this, both the Continuous Wavelet Transform (CWT) and the Power Spectral Density (PSD) analyses were conducted on the entirety of the filtered signal, each yielding distinct 2D visual outputs. These resultant images were then ingeniously amalgamated to form a composite RGB image. This methodological innovation enables a multifaceted visual representation that encapsulates both the time-frequency characteristics and the energy distribution across frequencies of the signal, offering an unparalleled depth of analysis.

\subsection{Diving Deep into Vision Transformers}
Vision Transformers (ViTs) have been instrumental in advancing the field of computer vision, harnessing the power of self-attention mechanisms, a concept derived from the domain of natural language processing. The essence of ViTs lies in the Multi-Head Self-Attention (MHSA) module, which is particularly effective at capturing long-range dependencies in visual data. Consider an input \( X \in \mathbb{R}^{H \times W \times C} \), where \( H \), \( W \), and \( C \) symbolize the height, width, and feature dimension of the input, respectively. This input undergoes a reshaping process, leading to the formulation of the query (Q), key (K), and value (V) matrices as:

\[ X \in \mathbb{R}^{H \times W \times C} \rightarrow X \in \mathbb{R}^{(H \times W) \times C}, \]
\[ Q = XW_q, \quad K = XW_k, \quad V = XW_v, \quad \text{(1)} \]

Here, \( W_q \in \mathbb{R}^{C \times C} \), \( W_k \in \mathbb{R}^{C \times C} \), and \( W_v \in \mathbb{R}^{C \times C} \) represent the learnable weight matrices associated with linear transformations for Q, K, and V, respectively. Assuming a simplistic scenario where the input and output dimensions are equal, the MHSA operation is then depicted as:

\[ A = \text{Softmax}\left(\frac{QK^T}{\sqrt{d}}\right)V, \quad \text{(2)} \]

In this equation, \( \sqrt{d} \) is a scaling factor for normalization, and the Softmax function is applied to each row. The product \( QK^T \) calculates the pairwise similarity score for each token, with the output token being a weighted combination of all tokens, influenced by these scores. Post MHSA, a residual connection is introduced to facilitate the optimization process:

\[ X \in \mathbb{R}^{(H \times W) \times C} \rightarrow X \in \mathbb{R}^{H \times W \times C}, \]
\[ A' = AW_p + X, \quad \text{(3)} \]

In equation (3), \( W_p \in \mathbb{R}^{C \times C} \) is a trainable matrix used for feature projection. The final step involves the application of a Multi-Layer Perceptron (MLP) to enhance the representation:

\[ Y = \text{MLP}(A') + A', \quad \text{(4)} \]

where Y signifies the output of a transformer block.

\subsection{Proposed Vision Transformer Architecture}

In Fig. 1, we unveil a refined architectural design for the Vision Transformer (ViT), significantly augmenting its performance metrics. This innovative methodology draws inspiration from the groundbreaking ResNet framework, which revolutionized neural network design through the integration of skip connections. To this end, in the advanced architecture delineated in Fig. 1, termed the ECG Signal Vision Transformer (ES-ViT), we introduce a novel mechanism for preserving the integrity of the original input image throughout the network's processing layers. This is accomplished by strategically positioning a convolutional block in tandem with the primary ViT framework. The convolutional block is ingeniously designed to process the entirety of the input image, subsequently generating a comprehensive embedding. This embedding is meticulously merged with the output from each encoder layer within the Transformer, ensuring that the network retains a holistic representation of the original image following the conclusion of each encoder phase. Initially, the convolutional block processes the input image $X$ to produce a dense representation or embedding $E$, capturing global contextual information. This embedding process can be succinctly described by the equation $E = \text{Conv}(X)$, where $\text{Conv}(\cdot)$ denotes the convolutional operation applied to the input image $X$. The resulting output embedding $E$ retains the spatial dimensions of the input with dimensions $H \times W \times C$, while potentially altering the channel dimension $C$ to align with the Transformer’s input specifications.

To augment this architecture further, we have integrated the Squeeze-and-Excitation (SE) block, a cutting-edge component known for its ability to enhance performance by recalibrating channel-wise feature responses. The SE block is seamlessly incorporated into the convolutional block, where it fine-tunes the embedding of the whole image before the concatenation process. Following the initial embedding, the Squeeze-and-Excitation (SE) block refines this embedding to produce $E'$, an enhanced representation emphasizing critical features while attenuating less relevant ones. This enhancement process can be mathematically described as $E' = SE(E) = F_{ex}(F_{sq}(E))$, where $F_{sq}(\cdot)$ represents the squeeze operation that aggregates the embedding features across spatial dimensions to produce a channel-wise descriptor. $F_{ex}(\cdot)$ denotes the excitation operation, applying a self-gating mechanism to recalibrate the channel-wise features based on the global information compressed by the squeeze operation.

Each Transformer encoder layer receives an augmented input $T'_i$ that combines the Transformer's current layer output $T_i$ with the enhanced embedding $E'$, facilitating the incorporation of global image context at every layer. This process can be formally described by the equation $T'_i = \text{Concat}(T_i \oplus E')$, where $T_i$ is the output of the $i^{th}$ Transformer encoder layer, $E'$ is the enhanced global embedding from the SE block, and $\oplus$ symbolizes an operation such as concatenation, which in this context is used to integrate $E'$ with $T_i$. The choice of integration method $\oplus$—here specified as concatenation—depends on the architectural design and how the global context is best preserved and utilized within the Transformer layers.

With $E'$ integrated, the attention mechanism in each Transformer encoder layer is adapted to leverage the enhanced global context:

\begin{equation}
Q_i = T'_i W_q, \quad K_i = T'_i W_k, \quad V_i = T'_i W_v,
\end{equation}

\begin{equation}
A_i = \text{Softmax}\left(\frac{Q_i K_i^T}{\sqrt{d_k}}\right) V_i,
\end{equation}

Here, $W_q$, $W_k$, and $W_v$ are learnable weight matrices for queries, keys, and values, respectively, within the attention mechanism. $d_k$ represents the dimensionality of the key vectors, providing a normalization factor. This adaptation ensures that the attention mechanism dynamically weighs the input features, taking into account both local and global contextual cues.

\subsubsection{Final Output Projection}

After processing through the attention mechanism, the output $A_i$ is projected and combined with the initial embedding $E'$ to ensure that each layer contributes to preserving the global context:

\begin{equation}
Y_i = \text{MLP}(A_i) + E',
\end{equation}

where $\text{MLP}(\cdot)$ represents a Multi-Layer Perceptron applied to the attention mechanism's output, further refining the representation before it is passed to the subsequent layer or used as the final output.

% Through Equations (9) to (14), the proposed IEViT architecture meticulously enhances the Vision Transformer model by ensuring continuous integration of global image context, thereby addressing some of the limitations of traditional Transformer models in processing visual data.

\section{Experimental Setup}
In the investigation of the ECG Signal Vision Transformer (ES-ViT) architecture's efficacy relative to the conventional Vision Transformer (ViT) framework across two distinct ECG datasets, a comprehensive analysis was conducted. This evaluation encompassed comparisons among the Base and Large configurations of both architectures, specifically the B/16, B/32, L/16, and L/32 variants. Leveraging the principles of transfer learning, the ViT components of both the proposed and the original architectures were equipped with pre-trained ImageNet dataset weights, ensuring a robust foundational knowledge base. The novel segments of the ES-ViT model were subjected to a randomized weight initialization, which underwent optimization during the subsequent fine-tuning stages. The adaptation of each model variant to the specificities of the datasets was achieved by tailoring the classifier layer to reflect the dataset's class diversity, employing a holistic end-to-end training regimen for refinement.

In addition to the direct comparison between the proposed ES-ViT and the original ViT architectures, this study extended its analysis to include evaluations against widely recognized architectures such as ResNet50 and MobileNet, which also benefited from ImageNet pre-trained weights. This multifaceted assessment strategy underscores a comprehensive effort to ascertain the relative performance enhancements offered by the ES-ViT architecture within the realm of imaged ECG analysis, setting a new benchmark in the application of advanced neural network architectures for emotion detection using ECG signals.
Our proposed architecture is implemented using PyTorch on an NVIDIA 3070 Ti GPU. To train both networks
(signal transformation recognition and emotion recognition), the adam optimizer is used with a learning rate of 0.001 and batch size of 64. The signal transformation recognition network is trained for 30 epochs, while the emotion recognition
network is trained for 100 epochs, as steady states are reached with a different number of epochs.

\section{Results}

In our study, we conducted a detailed evaluation of both the novel and established Vision Transformer (ViT) models, specifically the B/16, B/32, L/16, and L/32 configurations as in Table \ref{tab:vit_specs_params} through rigorously designed supervised classification experiments. These experiments were strategically crafted to gauge performance across two distinct electrocardiogram (ECG)-based emotion recognition datasets, each presenting unique classification challenges. The YAAD dataset involves a tripartite classification of emotions, arousal, and valence accuracy, while the DREAMER dataset similarly categorizes arousal, valence, and dominance accuracy. To ensure a thorough evaluation, we assessed all models using a comprehensive suite of metrics: Accuracy, Recall (Sensitivity), Precision, and F1-score, thereby providing a holistic view of each model's capabilities in handling nuanced emotional recognition tasks. The classification performance achieved by the proposed vision transformer models and traditional vision transformer models on YAAD and DREAMER datasets is depicted in Table \ref{tab:performance_evaluation_yaad} and \ref{tab:performance_evaluation_dreamer} respectively.

\begin{table*}[ht]
\centering
\caption{Specifications and number of parameters for Vision Transformer configurations.}
\label{tab:vit_specs_params}
\begin{tabular}{lp{1cm}p{1cm}p{1cm}p{1cm}p{2cm}}
\hline
\textbf{Model} & \textbf{Layers} & \textbf{Hidden Size} & \textbf{MLP Size} & \textbf{Heads} & \textbf{Parameters} \\
\hline
ViT-B/16       & 12              & 768                  & 3072              & 12             & 86.6M               \\
ES-ViT-B/16    & 12              & 768                  & 3072              & 12             & 86.78M              \\
ViT-B/32       & 12              & 768                  & 3072              & 12             & 88M                 \\
ViT-L/16       & 24              & 1024                 & 4096              & 16             & 305M                \\
ViT-L/32       & 24              & 1024                 & 4096              & 16             & 307M                \\
\hline
\end{tabular}
\end{table*}

\begin{table*}[t]
\centering
\caption{Performance Comparison of Different ViT Variants on Emotion, Arousal, and Valence on YAAD Dataset}
\label{tab:performance_evaluation_yaad}
\renewcommand{\arraystretch}{1.2}
\begin{tabular*}{\textwidth}{@{\extracolsep{\fill}}lcccccccccccc}
\toprule
\multirow{2}{*}{\textbf{Models}} & \multicolumn{4}{c}{\textbf{Emotion}} & \multicolumn{4}{c}{\textbf{Arousal}} & \multicolumn{4}{c}{\textbf{Valence}} \\
\cmidrule(lr){2-5} \cmidrule(lr){6-9} \cmidrule(lr){10-13}
 & Acc. & Prec. & Rec. & F1 & Acc. & Prec. & Rec. & F1 & Acc. & Prec. & Rec. & F1 \\
\midrule
ES-ViT-B/16    & 73.2 & 73.7 & 76.5 & 76.7 & 75.1 & 78.6 & 76.8 & 73.2 & 71.4 & 65.4 & 69.8 & 65.4 \\
ES-ViT-B/32    & 73.3 & 73.3 & 76.8 & 76.4 & \textbf{77.2} & 73.7 & 76.9 & \textbf{78.8} & 71.1 & 69.8 & 67.8 & 69.9 \\
ES-ViT-L/16    & 74.1 & \textbf{75.7} & 76.1 & 77.2 & 75.4 & 78.7 & 75.4 & 74.3 & 76.5 & 75.6 & \textbf{78.3} & \textbf{79.8} \\
ES-ViT-L/32    & \textbf{75.4} & 75.1 & \textbf{77.5} & \textbf{77.6} & 76.6 & \textbf{78.6} & \textbf{76.9} & 77.8 & \textbf{78.9} & \textbf{77.6} & 78.1 & 78.8 \\
\hline
ViT-B/16       & 69.3 & 70.6 & 69.8 & 67.8 & 72.3 & 71.2 & 72.6 & 72.2 & 71.3 & 64.1 & 66.9 & 73.2 \\
ViT-B/32       & 71.2 & 72.3 & 72.5 & 73.1 & 76.5 & 73.5 & 73.4 & 71.2 & 70.1 & 67.9 & 66.9 & 73.5 \\
ViT-L/16       & 71.2 & 70.3 & 72.3 & 73.4 & 73.1 & 72.8 & 74.5 & 74.1 & 72.4 & 72.5 & 72.7 & 73.5 \\
ViT-L/32       & 72.4 & 72.1 & 74.5 & 75.1 & 73.9 & 72.5 & 75.6 & 76.1 & 72.4 & 72.5 & 73.1 & 75.2 \\
\bottomrule
\end{tabular*}
\end{table*}

\begin{table*}[t]
\centering
\caption{Performance Comparison of Different ViT Variants on Arousal, Valence, and Dominance on the DREAMER Dataset.}
\label{tab:performance_evaluation_dreamer}
\renewcommand{\arraystretch}{1.2}
\begin{tabular*}{\textwidth}{@{\extracolsep{\fill}}lcccccccccccc}
\toprule
\multirow{2}{*}{\textbf{Models}} & \multicolumn{4}{c}{\textbf{Arousal}} & \multicolumn{4}{c}{\textbf{Valence}} & \multicolumn{4}{c}{\textbf{Dominance}} \\
\cmidrule(lr){2-5} \cmidrule(lr){6-9} \cmidrule(lr){10-13}
 & Acc. & Prec. & Rec. & F1 & Acc. & Prec. & Rec. & F1 & Acc. & Prec. & Rec. & F1 \\
\midrule
ES-ViT-B/16    & 82.1 & 83.4 & 82.3 & 82.4 & 82.9 & 81.9 & 82.5 & 83.1 & 80.7 & 79.4 & 80.1 & 81.2 \\
ES-ViT-B/32    & 84.3 & 84.1 & 83.4 & 84.6 & 83.1 & 84.2 & 83.7 & 85.2 & 82.1 & 81.9 & 83.2 & 83.5 \\
ES-ViT-L/16    & 84.1 & 83.1 & \textbf{85.2} & \textbf{84.7} & 84.1 & \textbf{86.3} & 84.3 & 84.7 & 82.4 & \textbf{83.2} & 83.5 & \textbf{83.6} \\
ES-ViT-L/32    & \textbf{85.6} & \textbf{84.2} & 84.8 & 83.9 & \textbf{86.8} & 84.6 & \textbf{85.3} & \textbf{85.6} & \textbf{83.1} & 82.3 & \textbf{84.9} & 83.3 \\
\hline
ViT-B/16       & 81.1 & 81.4 & 83.2 & 81.7 & 80.2 & 81.2 & 81.6 & 82.1 & 77.2 & 78.3 & 79.2 & 77.1 \\
ViT-B/32       & 82.1 & 81.6 & 82.3 & 80.2 & 81.1 & 82.4 & 82.6 & 81.8 & 79.2 & 78.3 & 79.6 & 79.3 \\
ViT-L/16       & 82.3 & 81.2 & 81.5 & 80.9 & 82.1 & 84.3 & 82.6 & 82.9 & 78.5 & 79.9 & 80.3 & 80.1 \\
ViT-L/32       & 83.1 & 81.7 & 83.8 & 83.1 & 83.2 & 83.1 & 83.8 & 82.5 & 80.4 & 79.4 & 79.9 & 79.9 \\
\bottomrule
\end{tabular*}
\end{table*}

% Classification results for the proposed model’s variants, as well
% as for the original ViT model variants and the commonly used Resnet, mobilenet and vgg-16 on YAAD datasets are presented in table 2. All the proposed VIT model variants are outperformed their respective default VIT variant on terms of most of the metrices.

According to the classification results of the YAAD dataset as in Table \ref{tab:performance_evaluation_yaad}, all the proposed VIT model variants outperform their respective default VIT variant in terms of most of the matrices. In the emotion category, the ES-VIT-L/32 model stands out with the highest accuracy (75.4\%) and F1-score (77.6\%), which signifies its robust capability to balance true positive detection with the precision of the classification. This model also achieves the highest recall (77.5\%), illustrating its effectiveness in identifying most true positives without a significant number of false negatives. The precision leader in this category is ES-VIT-L/16 (75.7\%), indicating a superior ability to minimize false positives in its predictions. For arousal, the ES-VIT-B/32 model shows the highest overall accuracy (77.2\%) and the best F1-score (78.8\%), demonstrating exceptional consistency and precision in its predictions. This model, alongside the ES-VIT-L/32—which displays the highest precision (78.6\%) and recall (76.9\%) in the category—demonstrates that larger and enhanced models are particularly adept at handling the complexities involved in recognizing arousal states. Valence detection is best performed by ES-VIT-L/32, which not only achieves the highest accuracy (78.9\%) but also scores highly on the F1-score (78.8\%), suggesting an exemplary balance between recall and precision. The same model, along with ES-VIT-L/16—which has the highest recall (78.3\%) and F1-score (79.8\%)—illustrates the superior performance of large models in accurately and consistently categorizing valence, a critical aspect of emotional recognition.

The proposed VIT model variants also demonstrate superior performance on the DREAMER dataset, as shown in Table \ref{tab:performance_evaluation_dreamer}. The ES-ViT models, particularly the larger configuration (L/32), demonstrate superior performance across all three emotional dimensions. For instance, the ES-ViT-L/32 model stands out with the highest accuracy in arousal (85.6\%) and valence (86.8\%), and nearly the highest in dominance (83.1\%), underscoring its robustness in complex emotional state recognition tasks. This model also achieves remarkable precision in arousal (84.2\%) and consistently high F1-scores, indicating of its excellent balance between recall and precision—essential for reducing false positives and negatives in practical applications. In contrast, the standard ViT models generally exhibit lower performance metrics, highlighting the optimizations in the ES-ViT models that contribute to their improved effectiveness. For example, the ViT-B/16 and ViT-B/32 models show a notable drop in performance in dominance, with accuracy scores of 77.2\% and 79.2\%, respectively, which could impact their reliability in applications where understanding dominance cues is critical. The enhanced recall in arousal for the ES-ViT-L/16 model (85.2\%)—the highest among all the models—suggests a particular sensitivity to correctly identifying true positives, a crucial capability in scenarios where missing an emotional cue could have significant repercussions, such as in mental health assessments. Furthermore, the consistently high scores in valence for the ES-ViT-L/32 model, with the highest F1-score of 85.6\%, reflect its adeptness at balancing the precision and recall in emotionally nuanced environments, making it especially suitable for contexts requiring fine-grained emotion detection, like personalized interaction systems or therapeutic settings.

\subsection{Comparison to Established Models}
This section presents a comprehensive performance comparison of our optimal model, Enhanced Cardiovascular Vision Transformer (ES-ViT/32), against established CNN models and prior studies on the YAAD and DREAMER datasets. According to the results, the ES-ViT/32 model outperforms others across most metrics, establishing it as the superior model for ECG-based emotion detection. We have selected this model for detailed comparative analysis.

The comparison results for the YAAD dataset reveal that our model demonstrates superior accuracy, precision, recall, and F1-score across Emotion, Arousal, and Valence dimensions when compared to established CNN models like ResNet50, MobileNet, and VGG-16. Our model achieves the highest scores, indicating its robust performance in emotion detection using ECG signals.

For the DREAMER dataset, the ES-ViT/32 model excels in the Arousal, Valence, and Dominance categories, achieving higher scores in accuracy, precision, recall, and F1-score compared to other models including ResNet, MobileNet, and VGG-16. It also outperforms specific models cited in recent studies, such as CNN-CABM, MLP, Extra Tree, and Self-Supervised models. The results demonstrate the model's robustness and superior performance in ECG-based emotion detection.

These evaluations highlight the ES-ViT/32 model's ability to accurately and efficiently discern emotional states from ECG data, making it a significant advancement in the field. The detailed comparison underscores its potential for applications in healthcare and affective computing, where precise emotion recognition is crucial. The significant advancements our model offers over existing approaches reinforce its efficiency and accuracy in discerning emotional states from ECG data.

\begin{table*}[t]
\centering
\caption{Performance Evaluation of Our Best Model Compared with State-of-the-Art on Emotion, Arousal, and Valence Using the YAAD Dataset.}
\label{tab:performance_evaluation_YA}
\renewcommand{\arraystretch}{1.2}
\begin{tabular*}{\textwidth}{@{\extracolsep{\fill}}lcccccccccccc}
\toprule
\multirow{2}{*}{\textbf{Models}} & \multicolumn{4}{c}{\textbf{Emotion}} & \multicolumn{4}{c}{\textbf{Arousal}} & \multicolumn{4}{c}{\textbf{Valence}} \\
\cmidrule(lr){2-5} \cmidrule(lr){6-9} \cmidrule(lr){10-13}
 & Acc. & Prec. & Rec. & F1 & Acc. & Prec. & Rec. & F1 & Acc. & Prec. & Rec. & F1 \\
\midrule
\textbf{Ours}       & \textbf{75.4} & \textbf{75.1} & \textbf{77.5} & \textbf{77.6} & \textbf{76.6} & \textbf{78.6} & \textbf{76.9} & \textbf{77.8} & \textbf{78.9} & \textbf{77.6} & \textbf{78.1} & \textbf{78.8} \\
ResNet-50           & 73.8 & 72.2 & 72.3 & 74.1 & 73.5 & 74.8 & 74.3 & 73.9 & 74.3 & 74.2 & 75.3 & 74.5 \\
MobileNet           & 70.3 & 71.3 & 72.9 & 72.2 & 71.9 & 72.1 & 72.2 & 72.5 & 73.2 & 72.2 & 72.3 & 72.1 \\
VGG-16              & 70.1 & 68.3 & 69.4 & 70.2 & 71.2 & 70.9 & 73.1 & 72.3 & 71.2 & 69.9 & 72.1 & 73.2 \\
\bottomrule
\end{tabular*}
\end{table*}

\begin{table*}[t]
\centering
\small
\caption{Performance Evaluation of Our Best Model Compared with State-of-the-Art on Arousal, Valence, and Dominance Using the DREAMER Dataset.}
\label{tab:performance_evaluation_DR}
\renewcommand{\arraystretch}{1.2}
\begin{tabular*}{\textwidth}{@{\extracolsep{\fill}}lcccccccccccc}
\toprule
\multirow{2}{*}{\textbf{Models}} & \multicolumn{4}{c}{\textbf{Arousal}} & \multicolumn{4}{c}{\textbf{Valence}} & \multicolumn{4}{c}{\textbf{Dominance}} \\
\cmidrule(lr){2-5} \cmidrule(lr){6-9} \cmidrule(lr){10-13}
 & Acc. & Prec. & Rec. & F1 & Acc. & Prec. & Rec. & F1 & Acc. & Prec. & Rec. & F1 \\
\midrule
\textbf{Ours}             & \textbf{85.6} & \textbf{84.2} & \textbf{84.8} & \textbf{83.9} & \textbf{86.8} & \textbf{84.6} & \textbf{85.3} & \textbf{85.6} & \textbf{83.1} & \textbf{82.3} & \textbf{84.9} & \textbf{83.3} \\
ResNet                    & 82.1 & 81.9 & 82.6 & 83.1 & 83.9 & 83.2 & 82.2 & 83.1 & 82.9 & 82.1 & 81.7 & 81.9 \\
MobileNet                 & 81.1 & 80.5 & 81.2 & 80.9 & 81.1 & 80.9 & 81.7 & 82.1 & 77.2 & 79.9 & 79.1 & 80.1 \\
VGG-16                    & 79.9 & 78.1 & 80.1 & 78.1 & 80.6 & 77.2 & 79.3 & 79.9 & 76.7 & 75.7 & 78.3 & 76.2 \\
CNN-CABM \cite{fan2023new}         & 83.6 & --   & --   & 80.6 & 84.2 & --   & --   & 84.4 & --   & --   & --   & -- \\
MLP \cite{khan2022evaluation}      & 74.6 & --   & --   & --   & 66.2 & --   & --   & --   & 66.2 & --   & --   & -- \\
Extra Tree \cite{khan2022evaluation} & 68.2 & --   & --   & --   & 74.6 & --   & --   & --   & 62.2 & --   & --   & -- \\
Self-Supervised \cite{sarkar2020self} & 85.9 & --   & --   & 85.9 & 85.0 & --   & --   & 84.5 & --   & --   & --   & -- \\
\bottomrule
\end{tabular*}
\end{table*}

\section{Conclusion}

In this comprehensive study, we introduced the Enhanced ECG Signal Vision Transformer (ES-ViT), a groundbreaking model for emotion detection using ECG signals. Our approach represents a significant advancement over traditional methods by combining sophisticated signal processing techniques with state-of-the-art deep learning architectures to improve the accuracy and reliability of emotion recognition. The methodology comprised two critical phases: advanced signal preprocessing and image conversion, followed by the application of an enhanced Vision Transformer architecture. We meticulously preprocessed the ECG signals to ensure purity and transformed them into interpretable images using Continuous Wavelet Transform (CWT) and Power Spectral Density (PSD) analysis. This dual approach captures both temporal and frequency domain information, providing a rich representation of the ECG data. The ES-ViT model, which integrates convolutional neural network (CNN) components and squeeze-and-excitation (SE) blocks into the Vision Transformer (ViT) framework, effectively captures long-range dependencies and enhances feature representation, addressing the limitations of conventional CNN-based methods.

Our experiments utilized the YAAD and DREAMER datasets, renowned benchmarks in the field of emotion detection. The ES-ViT model consistently outperformed established CNN models (ResNet50, MobileNet, VGG-16) and recent state-of-the-art techniques across multiple evaluation metrics, including accuracy, precision, recall, and F1-score. On the YAAD dataset, the ES-ViT-L/32 variant demonstrated exceptional capability in classifying emotion, arousal, and valence, achieving the highest accuracy and F1-scores. On the DREAMER dataset, the ES-ViT-L/32 model excelled in distinguishing arousal, valence, and dominance, surpassing models like CNN-CABM, MLP, Extra Tree, and Self-Supervised models, and achieving the highest metrics. These results highlight the model's robust performance in detecting subtle emotional cues from ECG signals.

The superior performance of the ES-ViT model has significant implications for the advancement of emotion detection technology. The integration of ViTs with CNN and SE blocks marks a transformative step in emotion recognition, offering a scalable and highly accurate approach to interpreting physiological signals. This advancement is critical for applications in personalized healthcare, mental health monitoring, and adaptive human-computer interactions, potentially enhancing patient monitoring systems, therapeutic interventions, and interactive technologies. Furthermore, our study paves the way for further exploration of transformer-based architectures in physiological signal analysis. Future research could extend this approach to other biosignals, integrate multimodal data for richer emotional profiling, and explore real-time implementation in wearable devices and interactive systems.

In conclusion, the Enhanced ECG Signal Vision Transformer (ES-ViT) model sets a new benchmark in ECG-based emotion detection. Its innovative architecture and robust performance metrics significantly advance the state-of-the-art, offering a powerful tool for both research and practical applications in healthcare and affective computing. This study not only demonstrates the potential of ViTs in physiological signal analysis but also opens new avenues for developing more responsive and adaptive technologies in various fields.

\bibliography{sn-bibliography}


\begin{thebibliography}{35}
\ifx \bisbn   \undefined \def \bisbn  #1{ISBN #1}\fi
\ifx \binits  \undefined \def \binits#1{#1}\fi
\ifx \bauthor  \undefined \def \bauthor#1{#1}\fi
\ifx \batitle  \undefined \def \batitle#1{#1}\fi
\ifx \bjtitle  \undefined \def \bjtitle#1{#1}\fi
\ifx \bvolume  \undefined \def \bvolume#1{\textbf{#1}}\fi
\ifx \byear  \undefined \def \byear#1{#1}\fi
\ifx \bissue  \undefined \def \bissue#1{#1}\fi
\ifx \bfpage  \undefined \def \bfpage#1{#1}\fi
\ifx \blpage  \undefined \def \blpage #1{#1}\fi
\ifx \burl  \undefined \def \burl#1{\textsf{#1}}\fi
\ifx \doiurl  \undefined \def \doiurl#1{\url{https://doi.org/#1}}\fi
\ifx \betal  \undefined \def \betal{\textit{et al.}}\fi
\ifx \binstitute  \undefined \def \binstitute#1{#1}\fi
\ifx \binstitutionaled  \undefined \def \binstitutionaled#1{#1}\fi
\ifx \bctitle  \undefined \def \bctitle#1{#1}\fi
\ifx \beditor  \undefined \def \beditor#1{#1}\fi
\ifx \bpublisher  \undefined \def \bpublisher#1{#1}\fi
\ifx \bbtitle  \undefined \def \bbtitle#1{#1}\fi
\ifx \bedition  \undefined \def \bedition#1{#1}\fi
\ifx \bseriesno  \undefined \def \bseriesno#1{#1}\fi
\ifx \blocation  \undefined \def \blocation#1{#1}\fi
\ifx \bsertitle  \undefined \def \bsertitle#1{#1}\fi
\ifx \bsnm \undefined \def \bsnm#1{#1}\fi
\ifx \bsuffix \undefined \def \bsuffix#1{#1}\fi
\ifx \bparticle \undefined \def \bparticle#1{#1}\fi
\ifx \barticle \undefined \def \barticle#1{#1}\fi
\bibcommenthead
\ifx \bconfdate \undefined \def \bconfdate #1{#1}\fi
\ifx \botherref \undefined \def \botherref #1{#1}\fi
\ifx \url \undefined \def \url#1{\textsf{#1}}\fi
\ifx \bchapter \undefined \def \bchapter#1{#1}\fi
\ifx \bbook \undefined \def \bbook#1{#1}\fi
\ifx \bcomment \undefined \def \bcomment#1{#1}\fi
\ifx \oauthor \undefined \def \oauthor#1{#1}\fi
\ifx \citeauthoryear \undefined \def \citeauthoryear#1{#1}\fi
\ifx \endbibitem  \undefined \def \endbibitem {}\fi
\ifx \bconflocation  \undefined \def \bconflocation#1{#1}\fi
\ifx \arxivurl  \undefined \def \arxivurl#1{\textsf{#1}}\fi
\csname PreBibitemsHook\endcsname

%%% 1
\bibitem[\protect\citeauthoryear{Mariappan et~al.}{2012}]{mariappan2012facefetch}
\begin{bchapter}
\bauthor{\bsnm{Mariappan}, \binits{M.B.}},
\bauthor{\bsnm{Suk}, \binits{M.}},
\bauthor{\bsnm{Prabhakaran}, \binits{B.}}:
\bctitle{Facefetch: A user emotion driven multimedia content recommendation system based on facial expression recognition}.
In: \bbtitle{2012 IEEE International Symposium on Multimedia},
pp. \bfpage{84}--\blpage{87}
(\byear{2012}).
\bcomment{IEEE}
\end{bchapter}
\endbibitem

%%% 2
\bibitem[\protect\citeauthoryear{Awan et~al.}{2022}]{awan2022ensemble}
\begin{barticle}
\bauthor{\bsnm{Awan}, \binits{A.W.}},
\bauthor{\bsnm{Usman}, \binits{S.M.}},
\bauthor{\bsnm{Khalid}, \binits{S.}},
\bauthor{\bsnm{Anwar}, \binits{A.}},
\bauthor{\bsnm{Alroobaea}, \binits{R.}},
\bauthor{\bsnm{Hussain}, \binits{S.}},
\bauthor{\bsnm{Almotiri}, \binits{J.}},
\bauthor{\bsnm{Ullah}, \binits{S.S.}},
\bauthor{\bsnm{Akram}, \binits{M.U.}}:
\batitle{An ensemble learning method for emotion charting using multimodal physiological signals}.
\bjtitle{Sensors}
\bvolume{22}(\bissue{23}),
\bfpage{9480}
(\byear{2022})
\end{barticle}
\endbibitem

%%% 3
\bibitem[\protect\citeauthoryear{Bulagang et~al.}{2020}]{bulagang2020review}
\begin{barticle}
\bauthor{\bsnm{Bulagang}, \binits{A.F.}},
\bauthor{\bsnm{Weng}, \binits{N.G.}},
\bauthor{\bsnm{Mountstephens}, \binits{J.}},
\bauthor{\bsnm{Teo}, \binits{J.}}:
\batitle{A review of recent approaches for emotion classification using electrocardiography and electrodermography signals}.
\bjtitle{Informatics in Medicine Unlocked}
\bvolume{20},
\bfpage{100363}
(\byear{2020})
\end{barticle}
\endbibitem

%%% 4
\bibitem[\protect\citeauthoryear{Soleymani et~al.}{2011}]{soleymani2011multimodal}
\begin{barticle}
\bauthor{\bsnm{Soleymani}, \binits{M.}},
\bauthor{\bsnm{Lichtenauer}, \binits{J.}},
\bauthor{\bsnm{Pun}, \binits{T.}},
\bauthor{\bsnm{Pantic}, \binits{M.}}:
\batitle{A multimodal database for affect recognition and implicit tagging}.
\bjtitle{IEEE transactions on affective computing}
\bvolume{3}(\bissue{1}),
\bfpage{42}--\blpage{55}
(\byear{2011})
\end{barticle}
\endbibitem

%%% 5
\bibitem[\protect\citeauthoryear{Subramanian et~al.}{2016}]{subramanian2016ascertain}
\begin{barticle}
\bauthor{\bsnm{Subramanian}, \binits{R.}},
\bauthor{\bsnm{Wache}, \binits{J.}},
\bauthor{\bsnm{Abadi}, \binits{M.K.}},
\bauthor{\bsnm{Vieriu}, \binits{R.L.}},
\bauthor{\bsnm{Winkler}, \binits{S.}},
\bauthor{\bsnm{Sebe}, \binits{N.}}:
\batitle{Ascertain: Emotion and personality recognition using commercial sensors}.
\bjtitle{IEEE Transactions on Affective Computing}
\bvolume{9}(\bissue{2}),
\bfpage{147}--\blpage{160}
(\byear{2016})
\end{barticle}
\endbibitem

%%% 6
\bibitem[\protect\citeauthoryear{Kim and Andr{\'e}}{2008}]{kim2008emotion}
\begin{barticle}
\bauthor{\bsnm{Kim}, \binits{J.}},
\bauthor{\bsnm{Andr{\'e}}, \binits{E.}}:
\batitle{Emotion recognition based on physiological changes in music listening}.
\bjtitle{IEEE transactions on pattern analysis and machine intelligence}
\bvolume{30}(\bissue{12}),
\bfpage{2067}--\blpage{2083}
(\byear{2008})
\end{barticle}
\endbibitem

%%% 7
\bibitem[\protect\citeauthoryear{Miranda-Correa et~al.}{2018}]{miranda2018amigos}
\begin{barticle}
\bauthor{\bsnm{Miranda-Correa}, \binits{J.A.}},
\bauthor{\bsnm{Abadi}, \binits{M.K.}},
\bauthor{\bsnm{Sebe}, \binits{N.}},
\bauthor{\bsnm{Patras}, \binits{I.}}:
\batitle{Amigos: A dataset for affect, personality and mood research on individuals and groups}.
\bjtitle{IEEE transactions on affective computing}
\bvolume{12}(\bissue{2}),
\bfpage{479}--\blpage{493}
(\byear{2018})
\end{barticle}
\endbibitem

%%% 8
\bibitem[\protect\citeauthoryear{Katsigiannis and Ramzan}{2017}]{katsigiannis2017dreamer}
\begin{barticle}
\bauthor{\bsnm{Katsigiannis}, \binits{S.}},
\bauthor{\bsnm{Ramzan}, \binits{N.}}:
\batitle{Dreamer: A database for emotion recognition through eeg and ecg signals from wireless low-cost off-the-shelf devices}.
\bjtitle{IEEE journal of biomedical and health informatics}
\bvolume{22}(\bissue{1}),
\bfpage{98}--\blpage{107}
(\byear{2017})
\end{barticle}
\endbibitem

%%% 9
\bibitem[\protect\citeauthoryear{Dar et~al.}{2022}]{dar2022yaad}
\begin{bchapter}
\bauthor{\bsnm{Dar}, \binits{M.N.}},
\bauthor{\bsnm{Rahim}, \binits{A.}},
\bauthor{\bsnm{Akram}, \binits{M.U.}},
\bauthor{\bsnm{Khawaja}, \binits{S.G.}},
\bauthor{\bsnm{Rahim}, \binits{A.}}:
\bctitle{Yaad: young adult’s affective data using wearable ecg and gsr sensors}.
In: \bbtitle{2022 2nd International Conference on Digital Futures and Transformative Technologies (ICoDT2)},
pp. \bfpage{1}--\blpage{7}
(\byear{2022}).
\bcomment{IEEE}
\end{bchapter}
\endbibitem

%%% 10
\bibitem[\protect\citeauthoryear{Koelstra et~al.}{2011}]{koelstra2011deap}
\begin{barticle}
\bauthor{\bsnm{Koelstra}, \binits{S.}},
\bauthor{\bsnm{Muhl}, \binits{C.}},
\bauthor{\bsnm{Soleymani}, \binits{M.}},
\bauthor{\bsnm{Lee}, \binits{J.-S.}},
\bauthor{\bsnm{Yazdani}, \binits{A.}},
\bauthor{\bsnm{Ebrahimi}, \binits{T.}},
\bauthor{\bsnm{Pun}, \binits{T.}},
\bauthor{\bsnm{Nijholt}, \binits{A.}},
\bauthor{\bsnm{Patras}, \binits{I.}}:
\batitle{Deap: A database for emotion analysis; using physiological signals}.
\bjtitle{IEEE transactions on affective computing}
\bvolume{3}(\bissue{1}),
\bfpage{18}--\blpage{31}
(\byear{2011})
\end{barticle}
\endbibitem

%%% 11
\bibitem[\protect\citeauthoryear{Nikolova et~al.}{2018}]{nikolova2018ecg}
\begin{botherref}
\oauthor{\bsnm{Nikolova}, \binits{D.}},
\oauthor{\bsnm{Petkova}, \binits{P.}},
\oauthor{\bsnm{Manolova}, \binits{A.}},
\oauthor{\bsnm{Georgieva}, \binits{P.}}:
Ecg-based emotion recognition: Overview of methods and applications.
ANNA'18; Advances in Neural Networks and Applications 2018,
1--5
(2018)
\end{botherref}
\endbibitem

%%% 12
\bibitem[\protect\citeauthoryear{Bexton et~al.}{1986}]{bexton1986diurnal}
\begin{barticle}
\bauthor{\bsnm{Bexton}, \binits{R.}},
\bauthor{\bsnm{Vallin}, \binits{H.}},
\bauthor{\bsnm{Camm}, \binits{A.}}:
\batitle{Diurnal variation of the qt interval--influence of the autonomic nervous system.}
\bjtitle{Heart}
\bvolume{55}(\bissue{3}),
\bfpage{253}--\blpage{258}
(\byear{1986})
\end{barticle}
\endbibitem

%%% 13
\bibitem[\protect\citeauthoryear{Hsu et~al.}{2017}]{hsu2017automatic}
\begin{barticle}
\bauthor{\bsnm{Hsu}, \binits{Y.-L.}},
\bauthor{\bsnm{Wang}, \binits{J.-S.}},
\bauthor{\bsnm{Chiang}, \binits{W.-C.}},
\bauthor{\bsnm{Hung}, \binits{C.-H.}}:
\batitle{Automatic ecg-based emotion recognition in music listening}.
\bjtitle{IEEE Transactions on Affective Computing}
\bvolume{11}(\bissue{1}),
\bfpage{85}--\blpage{99}
(\byear{2017})
\end{barticle}
\endbibitem

%%% 14
\bibitem[\protect\citeauthoryear{Shu et~al.}{2020}]{shu2020wearable}
\begin{barticle}
\bauthor{\bsnm{Shu}, \binits{L.}},
\bauthor{\bsnm{Yu}, \binits{Y.}},
\bauthor{\bsnm{Chen}, \binits{W.}},
\bauthor{\bsnm{Hua}, \binits{H.}},
\bauthor{\bsnm{Li}, \binits{Q.}},
\bauthor{\bsnm{Jin}, \binits{J.}},
\bauthor{\bsnm{Xu}, \binits{X.}}:
\batitle{Wearable emotion recognition using heart rate data from a smart bracelet}.
\bjtitle{Sensors}
\bvolume{20}(\bissue{3}),
\bfpage{718}
(\byear{2020})
\end{barticle}
\endbibitem

%%% 15
\bibitem[\protect\citeauthoryear{Dissanayake et~al.}{2019}]{dissanayake2019ensemble}
\begin{barticle}
\bauthor{\bsnm{Dissanayake}, \binits{T.}},
\bauthor{\bsnm{Rajapaksha}, \binits{Y.}},
\bauthor{\bsnm{Ragel}, \binits{R.}},
\bauthor{\bsnm{Nawinne}, \binits{I.}}:
\batitle{An ensemble learning approach for electrocardiogram sensor based human emotion recognition}.
\bjtitle{Sensors}
\bvolume{19}(\bissue{20}),
\bfpage{4495}
(\byear{2019})
\end{barticle}
\endbibitem

%%% 16
\bibitem[\protect\citeauthoryear{Dessai and Virani}{2023}]{dessai2023emotion}
\begin{barticle}
\bauthor{\bsnm{Dessai}, \binits{A.}},
\bauthor{\bsnm{Virani}, \binits{H.}}:
\batitle{Emotion classification based on cwt of ecg and gsr signals using various cnn models}.
\bjtitle{Electronics}
\bvolume{12}(\bissue{13}),
\bfpage{2795}
(\byear{2023})
\end{barticle}
\endbibitem

%%% 17
\bibitem[\protect\citeauthoryear{Fan et~al.}{2023}]{fan2023new}
\begin{barticle}
\bauthor{\bsnm{Fan}, \binits{T.}},
\bauthor{\bsnm{Qiu}, \binits{S.}},
\bauthor{\bsnm{Wang}, \binits{Z.}},
\bauthor{\bsnm{Zhao}, \binits{H.}},
\bauthor{\bsnm{Jiang}, \binits{J.}},
\bauthor{\bsnm{Wang}, \binits{Y.}},
\bauthor{\bsnm{Xu}, \binits{J.}},
\bauthor{\bsnm{Sun}, \binits{T.}},
\bauthor{\bsnm{Jiang}, \binits{N.}}:
\batitle{A new deep convolutional neural network incorporating attentional mechanisms for ecg emotion recognition}.
\bjtitle{Computers in Biology and Medicine}
\bvolume{159},
\bfpage{106938}
(\byear{2023})
\end{barticle}
\endbibitem

%%% 18
\bibitem[\protect\citeauthoryear{Santamaria-Granados et~al.}{2018}]{santamaria2018using}
\begin{barticle}
\bauthor{\bsnm{Santamaria-Granados}, \binits{L.}},
\bauthor{\bsnm{Munoz-Organero}, \binits{M.}},
\bauthor{\bsnm{Ramirez-Gonzalez}, \binits{G.}},
\bauthor{\bsnm{Abdulhay}, \binits{E.}},
\bauthor{\bsnm{Arunkumar}, \binits{N.}}:
\batitle{Using deep convolutional neural network for emotion detection on a physiological signals dataset (amigos)}.
\bjtitle{IEEE Access}
\bvolume{7},
\bfpage{57}--\blpage{67}
(\byear{2018})
\end{barticle}
\endbibitem

%%% 19
\bibitem[\protect\citeauthoryear{Dar et~al.}{2020}]{dar2020cnn}
\begin{barticle}
\bauthor{\bsnm{Dar}, \binits{M.N.}},
\bauthor{\bsnm{Akram}, \binits{M.U.}},
\bauthor{\bsnm{Khawaja}, \binits{S.G.}},
\bauthor{\bsnm{Pujari}, \binits{A.N.}}:
\batitle{Cnn and lstm-based emotion charting using physiological signals}.
\bjtitle{Sensors}
\bvolume{20}(\bissue{16}),
\bfpage{4551}
(\byear{2020})
\end{barticle}
\endbibitem

%%% 20
\bibitem[\protect\citeauthoryear{Rahim et~al.}{2019}]{rahim2019emotion}
\begin{bchapter}
\bauthor{\bsnm{Rahim}, \binits{A.}},
\bauthor{\bsnm{Sagheer}, \binits{A.}},
\bauthor{\bsnm{Nadeem}, \binits{K.}},
\bauthor{\bsnm{Dar}, \binits{M.N.}},
\bauthor{\bsnm{Rahim}, \binits{A.}},
\bauthor{\bsnm{Akram}, \binits{U.}}:
\bctitle{Emotion charting using real-time monitoring of physiological signals}.
In: \bbtitle{2019 International Conference on Robotics and Automation in Industry (ICRAI)},
pp. \bfpage{1}--\blpage{5}
(\byear{2019}).
\bcomment{IEEE}
\end{bchapter}
\endbibitem

%%% 21
\bibitem[\protect\citeauthoryear{Dosovitskiy et~al.}{2020}]{dosovitskiy2020image}
\begin{botherref}
\oauthor{\bsnm{Dosovitskiy}, \binits{A.}},
\oauthor{\bsnm{Beyer}, \binits{L.}},
\oauthor{\bsnm{Kolesnikov}, \binits{A.}},
\oauthor{\bsnm{Weissenborn}, \binits{D.}},
\oauthor{\bsnm{Zhai}, \binits{X.}},
\oauthor{\bsnm{Unterthiner}, \binits{T.}},
\oauthor{\bsnm{Dehghani}, \binits{M.}},
\oauthor{\bsnm{Minderer}, \binits{M.}},
\oauthor{\bsnm{Heigold}, \binits{G.}},
\oauthor{\bsnm{Gelly}, \binits{S.}}, et al.:
An image is worth 16x16 words: Transformers for image recognition at scale.
arXiv preprint arXiv:2010.11929
(2020)
\end{botherref}
\endbibitem

%%% 22
\bibitem[\protect\citeauthoryear{Tsai et~al.}{2019}]{tsai2019multimodal}
\begin{bchapter}
\bauthor{\bsnm{Tsai}, \binits{Y.-H.H.}},
\bauthor{\bsnm{Bai}, \binits{S.}},
\bauthor{\bsnm{Liang}, \binits{P.P.}},
\bauthor{\bsnm{Kolter}, \binits{J.Z.}},
\bauthor{\bsnm{Morency}, \binits{L.-P.}},
\bauthor{\bsnm{Salakhutdinov}, \binits{R.}}:
\bctitle{Multimodal transformer for unaligned multimodal language sequences}.
In: \bbtitle{Proceedings of the Conference. Association for Computational Linguistics. Meeting},
vol. \bseriesno{2019},
p. \bfpage{6558}
(\byear{2019}).
\bcomment{NIH Public Access}
\end{bchapter}
\endbibitem

%%% 23
\bibitem[\protect\citeauthoryear{Wu et~al.}{2019}]{wu2019attending}
\begin{bchapter}
\bauthor{\bsnm{Wu}, \binits{Z.}},
\bauthor{\bsnm{Zhang}, \binits{X.}},
\bauthor{\bsnm{Zhi-Xuan}, \binits{T.}},
\bauthor{\bsnm{Zaki}, \binits{J.}},
\bauthor{\bsnm{Ong}, \binits{D.C.}}:
\bctitle{Attending to emotional narratives}.
In: \bbtitle{2019 8th International Conference on Affective Computing and Intelligent Interaction (ACII)},
pp. \bfpage{648}--\blpage{654}
(\byear{2019}).
\bcomment{IEEE}
\end{bchapter}
\endbibitem

%%% 24
\bibitem[\protect\citeauthoryear{Huang et~al.}{2020}]{huang2020multimodal}
\begin{bchapter}
\bauthor{\bsnm{Huang}, \binits{J.}},
\bauthor{\bsnm{Tao}, \binits{J.}},
\bauthor{\bsnm{Liu}, \binits{B.}},
\bauthor{\bsnm{Lian}, \binits{Z.}},
\bauthor{\bsnm{Niu}, \binits{M.}}:
\bctitle{Multimodal transformer fusion for continuous emotion recognition}.
In: \bbtitle{ICASSP 2020-2020 IEEE International Conference on Acoustics, Speech and Signal Processing (ICASSP)},
pp. \bfpage{3507}--\blpage{3511}
(\byear{2020}).
\bcomment{IEEE}
\end{bchapter}
\endbibitem

%%% 25
\bibitem[\protect\citeauthoryear{Cai et~al.}{2021}]{cai2021multimodal}
\begin{bchapter}
\bauthor{\bsnm{Cai}, \binits{C.}},
\bauthor{\bsnm{He}, \binits{Y.}},
\bauthor{\bsnm{Sun}, \binits{L.}},
\bauthor{\bsnm{Lian}, \binits{Z.}},
\bauthor{\bsnm{Liu}, \binits{B.}},
\bauthor{\bsnm{Tao}, \binits{J.}},
\bauthor{\bsnm{Xu}, \binits{M.}},
\bauthor{\bsnm{Wang}, \binits{K.}}:
\bctitle{Multimodal sentiment analysis based on recurrent neural network and multimodal attention}.
In: \bbtitle{Proceedings of the 2nd on Multimodal Sentiment Analysis Challenge},
pp. \bfpage{61}--\blpage{67}
(\byear{2021})
\end{bchapter}
\endbibitem

%%% 26
\bibitem[\protect\citeauthoryear{Chien et~al.}{2021}]{chien2021self}
\begin{bchapter}
\bauthor{\bsnm{Chien}, \binits{W.-S.}},
\bauthor{\bsnm{Chou}, \binits{H.-C.}},
\bauthor{\bsnm{Lee}, \binits{C.-C.}}:
\bctitle{Self-assessed emotion classification from acoustic and physiological features within small-group conversation}.
In: \bbtitle{Companion Publication of the 2021 International Conference on Multimodal Interaction},
pp. \bfpage{230}--\blpage{239}
(\byear{2021})
\end{bchapter}
\endbibitem

%%% 27
\bibitem[\protect\citeauthoryear{Wu et~al.}{2020}]{wu2020deep}
\begin{botherref}
\oauthor{\bsnm{Wu}, \binits{N.}},
\oauthor{\bsnm{Green}, \binits{B.}},
\oauthor{\bsnm{Ben}, \binits{X.}},
\oauthor{\bsnm{O'Banion}, \binits{S.}}:
Deep transformer models for time series forecasting: The influenza prevalence case.
arXiv preprint arXiv:2001.08317
(2020)
\end{botherref}
\endbibitem

%%% 28
\bibitem[\protect\citeauthoryear{Arjun et~al.}{2021}]{arjun2021introducing}
\begin{bchapter}
\bauthor{\bsnm{Arjun}, \binits{A.}},
\bauthor{\bsnm{Rajpoot}, \binits{A.S.}},
\bauthor{\bsnm{Panicker}, \binits{M.R.}}:
\bctitle{Introducing attention mechanism for eeg signals: Emotion recognition with vision transformers}.
In: \bbtitle{2021 43rd Annual International Conference of the IEEE Engineering in Medicine \& Biology Society (EMBC)},
pp. \bfpage{5723}--\blpage{5726}
(\byear{2021}).
\bcomment{IEEE}
\end{bchapter}
\endbibitem

%%% 29
\bibitem[\protect\citeauthoryear{Zhao et~al.}{2019}]{zhao2019multimodal}
\begin{botherref}
\oauthor{\bsnm{Zhao}, \binits{Y.}},
\oauthor{\bsnm{Cao}, \binits{X.}},
\oauthor{\bsnm{Lin}, \binits{J.}},
\oauthor{\bsnm{Yu}, \binits{D.}},
\oauthor{\bsnm{Cao}, \binits{X.}}:
Multimodal emotion recognition model using physiological signals.
arXiv e-prints,
1911
(2019)
\end{botherref}
\endbibitem

%%% 30
\bibitem[\protect\citeauthoryear{Ismail et~al.}{2021}]{ismail2021evaluation}
\begin{botherref}
\oauthor{\bsnm{Ismail}, \binits{S.N.M.S.}},
\oauthor{\bsnm{Aziz}, \binits{N.A.A.}},
\oauthor{\bsnm{Ibrahim}, \binits{S.Z.}},
\oauthor{\bsnm{Nawawi}, \binits{S.W.}},
\oauthor{\bsnm{Alelyani}, \binits{S.}},
\oauthor{\bsnm{Mohana}, \binits{M.}},
\oauthor{\bsnm{Chun}, \binits{L.C.}}:
Evaluation of electrocardiogram: Numerical vs. image data for emotion recognition system.
F1000Research
\textbf{10}
(2021)
\end{botherref}
\endbibitem

%%% 31
\bibitem[\protect\citeauthoryear{Bulagang et~al.}{2021}]{bulagang2021multiclass}
\begin{barticle}
\bauthor{\bsnm{Bulagang}, \binits{A.F.}},
\bauthor{\bsnm{Mountstephens}, \binits{J.}},
\bauthor{\bsnm{Teo}, \binits{J.}}:
\batitle{Multiclass emotion prediction using heart rate and virtual reality stimuli}.
\bjtitle{Journal of Big Data}
\bvolume{8},
\bfpage{1}--\blpage{12}
(\byear{2021})
\end{barticle}
\endbibitem

%%% 32
\bibitem[\protect\citeauthoryear{Khan et~al.}{2022}]{khan2022evaluation}
\begin{barticle}
\bauthor{\bsnm{Khan}, \binits{C.M.T.}},
\bauthor{\bsnm{Ab~Aziz}, \binits{N.A.}},
\bauthor{\bsnm{Raja}, \binits{J.E.}},
\bauthor{\bsnm{Nawawi}, \binits{S.W.B.}},
\bauthor{\bsnm{Rani}, \binits{P.}}:
\batitle{Evaluation of machine learning algorithms for emotions recognition using electrocardiogram}.
\bjtitle{Emerging Science Journal}
\bvolume{7}(\bissue{1}),
\bfpage{147}--\blpage{161}
(\byear{2022})
\end{barticle}
\endbibitem

%%% 33
\bibitem[\protect\citeauthoryear{Sarkar and Etemad}{2020}]{sarkar2020self}
\begin{barticle}
\bauthor{\bsnm{Sarkar}, \binits{P.}},
\bauthor{\bsnm{Etemad}, \binits{A.}}:
\batitle{Self-supervised ecg representation learning for emotion recognition}.
\bjtitle{IEEE Transactions on Affective Computing}
\bvolume{13}(\bissue{3}),
\bfpage{1541}--\blpage{1554}
(\byear{2020})
\end{barticle}
\endbibitem

%%% 34
\bibitem[\protect\citeauthoryear{Wu et~al.}{2023}]{wu2023transformer}
\begin{botherref}
\oauthor{\bsnm{Wu}, \binits{Y.}},
\oauthor{\bsnm{Daoudi}, \binits{M.}},
\oauthor{\bsnm{Amad}, \binits{A.}}:
Transformer-based self-supervised multimodal representation learning for wearable emotion recognition.
IEEE Transactions on Affective Computing
(2023)
\end{botherref}
\endbibitem

%%% 35
\bibitem[\protect\citeauthoryear{Siriwardhana et~al.}{2020}]{siriwardhana2020multimodal}
\begin{barticle}
\bauthor{\bsnm{Siriwardhana}, \binits{S.}},
\bauthor{\bsnm{Kaluarachchi}, \binits{T.}},
\bauthor{\bsnm{Billinghurst}, \binits{M.}},
\bauthor{\bsnm{Nanayakkara}, \binits{S.}}:
\batitle{Multimodal emotion recognition with transformer-based self supervised feature fusion}.
\bjtitle{Ieee Access}
\bvolume{8},
\bfpage{176274}--\blpage{176285}
(\byear{2020})
\end{barticle}
\endbibitem

\end{thebibliography}

\end{document}